\RequirePackage{fix-cm}
\documentclass[twocolumn,epjc3]{svjour3}  
\smartqed  
\RequirePackage{graphicx}
\usepackage{graphicx}
\usepackage{amsmath, amssymb}
\usepackage{mathrsfs}
\usepackage{mathtools}
\usepackage{tensor}
\usepackage{cite}
\usepackage{leftindex} 
\usepackage{subcaption} 
\usepackage{float}
\usepackage{url}
\usepackage{mathptmx} 
\usepackage{color}

\allowdisplaybreaks

\journalname{Eur. Phys. J. C}
\begin{document}

\title{Cosmology in the De Donder-Weyl Formulation of Einstein-Cartan Gravity}


\author{Aarav Shah\thanksref{e1,addr1}
        \and
        Maxim Khlopov\thanksref{e2,addr2}
        \and
        Maxim Krasnov\thanksref{e3,addr3,addr4} 
}

\thankstext{e1}{e-mail: shahaarav103@zohomail.in}
\thankstext{e2}{email: khlopov@apc.in2p3.fr}
\thankstext{e3}{email:  morrowindman1@mail.ru}


\institute{International Center for Space and Cosmology, Ahmedabad University, Ahmedabad 380009, Gujarat, India  \label{addr1}
           \and
            Virtual Institute of Astroparticle Physics, 75018 Paris, France\label{addr2}
           \and
           National Research Nuclear University MEPhI, 115409 Moscow, Russia \label{addr3}
           \and
           Research Institute of Physics, Southern Federal University, 344090 Rostov-on-Don, Russia \label{addr4}     
}

\date{Received: date / Accepted: date}

\maketitle

\begin{abstract}
We investigate torsion-driven cosmological dynamics within the framework of Einstein-Cartan gravity using the De Donder-Weyl Hamiltonian formalism, where the tetrad and Lorentz connection act as independent variables and the Hamiltonian includes quadratic Riemann Cartan corrections. Embedding this theory in an FLRW background, we derive the corresponding torsion-modified Friedmann equations and analyze their solutions across radiation and matter-dominated epochs. The commonly assumed power law form $a(t)=\beta t^{\alpha}$ is shown to generate multiple solution branches, many of which can be considered to be “unphysical”. A hybrid solution, $a(t)=Ct^{\alpha}e^{Dt^{\beta}}$ emerges in the special case $g_1=0$, where the quadratic Riemann-Cartan term vanishes. For $g_1\not=0,$ the equations become nonlinear, precluding closed-form analytic solutions. These findings highlight the limitations of the power-law approximation and identify the restricted conditions under which torsion can coherently drive cosmic expansion.  
\end{abstract}

\section{Introduction}
General relativity (GR) \cite{Misner1973,Wald1984,Carroll2004,Wald1999_LRR,Will2014,PoissonWill2014,Padmanabhan2010,Sciama1962} remains the cornerstone of modern cosmology \cite{Baumann:2022cosmology,Mukhanov:2005fem, Weinberg:2008cosmology}, successfully describing gravitational dynamics across a vast range of scales. However, GR assumes a purely Riemannian spacetime, curved but torsion free. The Einstein-Cartan (EC) theory \cite{Cartan1922,Cartan1923,Cartan1924,Sciama1964a,Kibble1961,Trautman1972,Hehl1971,Hehl1973,Hehl1974,Hehl1976Foundations,Trautman1973,Hehl1979_book,Shapiro2002,HehlObukhov2007} generalizes this picture \footnote{There have been various extensions to General relativity, see for example \cite{Blagojevic2013,Kibble1961,Capozziello2015,shah2025ultraslowrollphasewarm}.} by admitting an antisymmetric component of the connection, representing torsion, which couples to the intrinsic spin of matter \cite{Hehl1976Foundations,Shapiro2002}. In such a Riemann-Cartan geometry, both curvature and torsion coexist, leading to a richer and a potentially more complete description of spacetime dynamics. 
\\
\\
Modern torsion cosmology \cite{Mavromatos2021,Hammond2002,HehlObukhov2007,Sezgin1981,Shie_2008,Poplawski2011_cc,Baekler2011,Smalley1978,Puetzfeld2008,Battista_2021,Battista2022,Battista_2022,Battista_2023,De_Falco_2023,De_Falco_2024} has two attractive phenomenological features. First, the intrinsic spin-torsion coupling in Einstein-Cartan theories produces a gravitational repulsion at very high densities which can avert the initial singularity and give a nonsingular bounce\cite{Poplawski2012,Bolejko_2020,Trautman1973}, providing an alternative to inflationary cosmology \footnote{See \cite{Poplawski2010} for more information on how Einstein-Cartan theories provide an alternative to inflationary behavior.}. Second, when torsion is promoted to a dynamical field (for example by including kinetic or quadratic curvature terms) it can produce late-time acceleration or mimic a cosmological constant like component without invoking an explicit $\Lambda$. These possibilities have been explored recently in quadratic gauge gravity and De Donder Weyl approaches \cite{DeDonder1930,Weyl1935,Caratheodory1929,GoldschmidtSternberg1973,Kastrup1983,Kanatchikov2000,Gotay1998,Forger2005,Struckmeier2008,Struckmeier2015,Vasak2020}, where torsion-driven dynamics may both replace $\Lambda$ and leave distinct imprints on the expansion history.  
\\
\\
These models often yield modified Friedmann equations, where torsion acts as an effective source term influencing the cosmic expansion rate \cite{KirschVasakStruckmeier2023,vanDeVennVasakKirschStruckmeier2022,Benisty2021,Shie_2008}. Earlier analyses typically focused on the radiation-dominated epoch, assuming that the scale factor follows a simple power-law evolution of the form $a(t)=\beta t^{\alpha}$. This assumption simplifies the field equations and provides analytic tractability, but it also implies different distinct branches of solutions, each of which have a different $\alpha$ and $C$.
\\
\\
In the present work, we revisit the cosmological implications of Einstein-Cartan gravity through the De Donder Weyl Hamiltonian formalism \cite{DeDonder1930,Weyl1935,Caratheodory1929,GoldschmidtSternberg1973,Kastrup1983,Kanatchikov2000,Gotay1998,Forger2005,Struckmeier2008,Struckmeier2015,Vasak2020}, a first-order, covariant approach that treats the tetrad and Lorentz connection as independent dynamical variables \cite{Vasak2023_CCGG}. This formalism introduces canonical momentum fields associated with spacetime geometry and leads naturally to a quadratic Riemann-Cartan Hamiltonian density, incorporating both curvature and torsion contributions. The resulting equations generalize Einstein’s field equations and give rise to a self-consistent framework for studying torsion-induced expansion.
\\
\\
We apply this formalism to a Friedmann Lema\^{i}tre Robertson Walker (FLRW) Universe \cite{Baumann:2022cosmology,Mukhanov:2005fem,Weinberg:2008cosmology}, embedding torsion into a homogeneous and isotropic background. Within this setting, we derive extended Friedmann equations that explicitly depend on the deformation parameters of the theory. We first recover known results for the radiation-dominated era \cite{vanDeVennVasakKirschStruckmeier2022}, but crucially extend the analysis to the matter-dominated epoch, where torsion modifies the scaling behavior of the scale factor in nontrivial ways. Our results demonstrate that, in addition to the conventional power-law solutions, the field equations admit a class of hybrid solutions of the form
\begin{equation}
    a(t)= Ct^{\alpha} exp(Kt^{\beta}),
\end{equation}
where $K$ is a function of the torsion of the Universe and vanishes in a torsion less Universe.
\\
\\
We find that the various branches of the pure-power law form $a(t)=\beta t^{\alpha}$ are highly unphysical, and pathological; inconsistent with observations like the critical density of a matter source during Big-Bang Nucleosynthesis (BBN) \cite{Cyburt2016,Fields2023,Grohs2023,PDG2024,Tytler2000}.
\\
\\
The paper is organized as follows. In Sec.\ref{Section 2}, we outline the De Donder–Weyl Hamiltonian formalism for Einstein–Cartan gravity and derive the corresponding field equations. Sec.\ref{Section 3} embeds this framework into a cosmological background, yielding the torsion-modified Friedmann equations and then goes on to study cosmological phases like  radiation domination (RDE) and matter domination (MDE). Finally, Sec.\ref{Section 4} summarizes the key results and discusses potential extensions toward early Universe dynamics and observational constraints.

\section{ Einstein–Cartan Gravity in the De Donder Weyl Framework}\label{Section 2}
We begin by formulating Einstein-Cartan theory within the De-Donder-Weyl-Hamiltonian formalism \cite{DeDonder1930,Weyl1935,Caratheodory1929,GoldschmidtSternberg1973,Kastrup1983,Kanatchikov2000,Gotay1998,Forger2005,Struckmeier2008,Struckmeier2015,Vasak2020}, a covariant first-order generalization of canonical dynamics to field theory originally developed by De Donder \cite{DeDonder1930} and Weyl \cite{Weyl1935}, and later expanded into modern covariant Hamiltonian and gauge formulations \cite{GoldschmidtSternberg1973,Kastrup1983,Kanatchikov2000,Gotay1998,Forger2005,Struckmeier2008,Struckmeier2015,Vasak2020}. In this framework, spacetime geometry is described by two independent dynamical variables- the Lorentz (or spin connection) $\omega^{i}_{j\alpha}$ and the tetrad field $e^{i}_{j\alpha}$- as in the Poincar\'{e} gauge treatments of gravity \cite{Kibble1961,Hehl1979_book,Blagojevic2013}. Greek indices refer to the coordinate frame and Latin indices to the local Lorentz frame. Throughout this work we adopt natural units $\hbar=c=1$ and the metric signature $(+,-,-,-)$.
\\
\\
Unlike the Arnowitt Deser Misner (ADM) decomposition of canonical GR \cite{Misner1973,Wald1984}, the DW approach treats all spacetime directions symmetrically and introduces poly-momenta, i.e tensorial generalizations of canonical momenta, conjugate to the derivatives of the dynamical fields \cite{Kastrup1983,Kanatchikov2000}. For gravity, the poly-momenta $k^{i}_{\mu\nu}$ and $q^{ij}_{\mu\nu}$ encode the “momentum content” of spacetime geometry, endowing curvature and torsion with independent kinetic degrees of freedom \cite{Kanatchikov2000,Struckmeier2015,KirschVasakStruckmeier2023}. This framework thus provides a natural setting to investigate the dynamical role of torsion in cosmology, extending earlier Einstein–Cartan analyses \cite{Cartan1922,Cartan1923,Cartan1924,Sciama1964a,Kibble1961,Trautman1972,Hehl1971,Hehl1973,Hehl1974,Hehl1976Foundations,Trautman1973,Hehl1979_book,Shapiro2002,HehlObukhov2007} and modern quadratic-gravity approaches where torsion can influence the cosmic expansion rate \cite{Poplawski2011_cc,Smalley1978,Benisty2021,Shie_2008}. 
\\
\\
The DW Hamiltonian density is constructed as a scalar function of the poly-momenta and the tetrad, extending the linear Einstein-Hilbert term by quadratic invariants built from curvature and torsion \cite{Kanatchikov2000,Vasak2020,Benisty2021}. A dimensionless coupling $g_1$ controls the strength of the quadratic Riemann-Cartan term, while $g_2,g_3,g_4$ parameterize geometric and matter couplings. The resulting formulation remains manifestly covariant and yields field equations that reduce to standard Einstein–Cartan gravity in the linear limit but include higher-order geometric corrections analogous to those appearing in quadratic gauge theories of gravity \cite{Shie_2008,Smalley1978,vanDeVennVasakKirschStruckmeier2022}.
\\
\\
In the DW approach one introduces an open cover of the spacetime manifold $M$ and a local orthonormal basis on each patch, spanned by a set of vierbeins (tetrads)
\begin{equation} \label{veirbiens}
    e_a=e_a^{\mu}\partial_{\mu} , a=1,...,4,
\end{equation}
which relate the coordinate and Lorentz frames \cite{Carroll2004,Trautman1972,Shapiro2002}. The metric tensor is then expressed through the tetrads as
\begin{equation}
    g_{\mu\nu}=e^{i}_{\mu}e^{\nu}_{\nu}\eta_{ij},
\end{equation}
where $\eta_{ij}$ denotes the Minkowski metric. This relation encodes how the local inertial structure of spacetime is projected into curved geometry \cite{Misner1973,Padmanabhan2010,Capozziello2015}.
\\
\\
The affine connection $\Gamma^{\lambda}_{\mu\nu}$ decomposes into the spin connection and the tetrad according to
\begin{equation}\label{Affine connection}
    \Gamma^{\lambda}_{\mu\nu}=-e^{i}_{\mu}\left(\frac{\partial e_{i}^{\lambda}}{\partial x^{\nu}}-\omega^{k}_{
    i\nu}e_{k}^{\lambda}\right),
\end{equation}
where the anti-symmetry of the spin connection, $\omega_{(ij)\alpha}$, ensures metric compatibility \cite{Sciama1964a,Trautman1973}.
Eqn.\eqref{Affine connection} reveals how local Lorentz rotations are coupled to coordinate changes.
\\
\\
The DW Hamiltonian (scalar) density, $H_{Gr}$, of spacetime dynamics extends the linear  Einstein-Hilbert ansatz by quadratic terms built from the momentum fields endowing the spacetime with kinetic energy and thus, inertia \cite{Vasak2020,Benisty2021}, fundamentally modifying its dynamics. We now set
\begin{equation}
    H=H_{Gr}+H_{matter},
\end{equation}
with a quadratic linear ansatz $H_{Gr}=H(q^2e^4,qe^2,k^2e^4,\epsilon)$. 
\\
\\
In explicit form \cite{KirschVasakStruckmeier2023,vanDeVennVasakKirschStruckmeier2022},
\begin{equation}\label{Hamiltonian}
\begin{split}
       H_{Gr}=\frac{1}{4g_1\epsilon}q_{l}^{m\alpha\beta}q_m^{l\xi \lambda}\eta_{kn}\eta_{ij}e^{k}_{\alpha}e^{n}_{\xi}e^{i}_{\beta}e^{j}_{\lambda}-g_2q_l^{m\alpha\beta}e^{l}_{\alpha}e^{n}_{\beta}\eta_{mn}\\+\frac{1}{2g_3\epsilon}k_{l}^{\alpha\beta}k_m^{\xi\lambda}\eta^{lm}\eta_{kn}\eta_{ij}\eta_{kn}\eta_{ij}e^{k}_{\alpha}e^{n}_{\xi}e^{i}_{\beta}e^{j}_{\lambda}+g_4\epsilon, 
\end{split}
\end{equation}
where $\epsilon=det(e^{i}_{\mu})=\sqrt{-g_{\mu\nu}}.$ $H_{matter}$ involves the coupling of matter fields to curved spacetime. The coupling constants $g_1$,$g_2$,$g_3$, and $g_4$ have dimensions $[g_1]=1,[g_2]=L^{-2},[g_3]=L^{-2}$ and $[g_4]=L^{-4}$ \cite{Benisty2021,Shie_2008}. Eqn.\eqref{Hamiltonian} shows that the first term, proportional to $g_1$, encodes the quadratic Riemann–Cartan contribution, while the remaining terms correspond to geometric, torsional, and vacuum energy contributions respectively. 
\\
\\
The gauging process then results in the action integral 
\begin{equation} \label{Action}
\begin{aligned}
S=\int_{V}d^4x L =\\ \int_{V} d^4x \frac{1}{2} \left[
\begin{aligned}
k_{i}^{\mu\nu}\left(\frac{\partial e^{i}_{\mu}}{\partial x^\nu}-\frac{\partial e^i_\nu}{\partial x^\mu}+\omega^{i}_{j\nu}e^j_{\mu}-\omega^{i}_{j\mu }\right)\\
+\frac{1}{2}q_{i}^{j\mu\nu}\left(\frac{\partial\omega^i_{j\mu}}{\partial x^\nu}-\frac{\partial\omega^i_{j\nu}}{\partial x^\mu}+\omega^{i}_{j\nu}e^j_{\mu}-\omega^{i}_{j\mu}e^{j}_{\nu} \right)\\
-H_{Gr}+L_{matter} 
\end{aligned}
\right],
\end{aligned}
\end{equation}
which serves as the starting point for deriving the canonical field equations of Einstein–Cartan gravity within the DW framework \cite{Vasak2020,vanDeVennVasakKirschStruckmeier2022}.
\\
\\
Variation of equation \eqref{Action} with respect to $q_i^{j\mu\nu}$ gives
\begin{equation} \label{First Canonical equation}
    \frac{\partial H_{Gr}}{\partial q_{i}^{j\mu\nu}}=\frac{1}{2}\left(\frac{\partial \omega^{i}_{j\mu}}{\partial x^\nu}-\frac{\partial \omega^{i}_{j\nu}}{\partial x^\mu}+\omega^i_{n\nu}\omega^n_{j\mu}-\omega^i_{n\mu}\omega^n_{j\nu}\right)=\frac{1}{2}R^i_{j\nu\mu},
\end{equation}
linking the poly-momenta $q^{i}_{j\mu\nu}$ to the the Riemann–Cartan curvature tensor. Curvature thus appears as a dynamical variable conjugate to the spin connection \cite{Kanatchikov2000,Vasak2020}.
\\
\\
Now, we will invoke torison as an additional structure to the underlying geometry. Such a structure extends the affine connection from the Christoffel symbol $\left\{^{\lambda}_{\mu\nu}\right\}
$ to
\begin{equation}\label{Expression 1}
    \Gamma^{\lambda}_{\mu\nu}=\left\{^{\lambda}_{\mu\nu}\right\}+K^{\lambda}_{\mu\nu},
\end{equation}
where the contortion tensor $K_{\lambda\mu\nu}$ is built from Cartan's torsion tensor \cite{Cartan1922,Cartan1923,Cartan1924,Sciama1964a,Kibble1961,Hehl1976Foundations}
\begin{equation} \label{Expression 2}
    S^{\lambda}_{\mu\nu}=\frac{1}{2}(\Gamma^{\lambda}_{\mu\nu}-\Gamma^{\lambda}_{\nu\mu}),
\end{equation}
and can be written as
\begin{equation} \label{Expression 3}
    K_{\lambda\mu\nu}=S_{\lambda\mu\nu}-S_{\lambda\nu\mu}+S_{\nu\mu\lambda}=-K_{\mu\lambda\nu}.
\end{equation}
These expressions introduce torsion explicitly as the antisymmetric part of the affine connection and relate it to the contortion tensor, which quantifies deviations from Levi-Civita geometry. 
\\
\\
From eqn.\eqref{Affine connection}, it follows that
\begin{equation}
    S^{i}_{\mu\nu}=\frac{\partial e^{i}_{\mu}}{\partial x^\nu}-\frac{\partial e^{i}_{\mu}}{\partial x^\nu}+\omega_{j\nu}^{i}e^{j}_{\mu}-\omega_{j\mu}^{i}e^{j}_{\nu}.
\end{equation}
Varying eqn.\eqref{Action} with respect to the field $k^{\mu\nu}_i$ then yields
\begin{equation} \label{Momentum Torsion}
    \frac{\partial H_{Gr}}{\partial k^{\mu\nu}_i}=\frac{1}{g_3\epsilon}k^i_{\mu\nu}=S^{i}_{\mu\nu},
\end{equation}
confirming that torsion acts as the canonical momentum of spacetime within this formalism,an interpretation emphasized in the covariant gauge gravity analyses of Struckmeier \cite{Struckmeier2015,Vasak2020,Struckmeier2008}. Furthermore, eqn.\eqref{Momentum Torsion} establishes that the poly-momentum $k^{i}_{\mu\nu}$ is directly proportional to the torsion tensor $S^{i}_{\mu\nu}$, confirming that torsion represents the canonical momentum of spacetime within this formalism.
\\
\\
Finally, the DW framework leads to a consistency (or zero energy) equation that extends Einstein’s field equations to include torsion and quadratic Riemann Cartan effects \cite{Benisty2021,Shie_2008}:
\begin{equation} \label{consti.eq.}
\Theta^{\nu}_{\mu}+T^{\nu}_{\mu}=0,
\end{equation}
with 
\begin{equation}\label{oMEGA}
    \Theta_{\mu}^{\nu}=\frac{1}{\sqrt{-g}}e^{i}_{\mu}\frac{\partial H_{Gr}}{\partial e^i_{\nu}},
\end{equation}
\begin{equation}
    T^{\nu}_{\mu}=\frac{1}{\sqrt{-g}}e^i_{\mu}\frac{\partial H_{matter}}{\partial e^{i}_{\nu}}.
\end{equation}
The “consistency equation” generalizes Einstein’s equations by including the energy–momentum tensor of spacetime geometry ($\Theta_{\mu}^{\nu}$) alongside that of matter ($T^{\nu}_{\mu}$), implementing a zero total energy condition.
\\
\\
In analogy to the energy momentum tensor of matter $T^{\mu\nu}$, we interpret $\Theta^{\mu\nu}$ as the energy momentum tensor of spacetime \cite{Vasak2020,Shapiro2002}. 
\\
\\
Calculating $\Theta^{\nu}_{\mu}$ using eqns.\eqref{Hamiltonian} and \eqref{oMEGA}, we have 
\begin{equation} \label{Space time geometry}
\begin{split}
        \Theta^{\mu\nu}=-g_1Q^{\mu\nu}+2g_1g_2(G^{\mu\nu}+3g_2g^{\mu\nu})\\+2g_3(S^{\xi\alpha\mu}S_{\xi\alpha}^{\nu}-\frac{1}{2}S^{\mu\alpha\beta}S_{\alpha\beta}^{\nu}-\frac{1}{4}g^{\mu\nu}S_{\chi\alpha\beta}S^{\chi\alpha\beta})+g_4g^{\mu\nu},
\end{split}
\end{equation}
where 
\begin{equation}
    G^{\mu\nu}=R^{(\mu\nu)}-\frac{1}{2}g^{\mu\nu}R,
\end{equation}
is the Einstein tensor and
\begin{equation}
Q^{\mu\nu}=R^{\alpha\beta\gamma\mu}R_{\alpha\beta\gamma}^{\nu}-\frac{1}{4}g^{\mu\nu}R^{\alpha\beta\gamma\xi}R_{\alpha\beta\gamma\xi},
\end{equation}
is the trace-free, (symmetric) quadratic  Riemann-Cartan concomitant \cite{Smalley1978,Benisty2021}. In eqn.\eqref{Space time geometry}, $\Theta_{\mu\nu}$ is expanded explicitly, showing separate contributions from the Einstein tensor, 
\\
quadratic Riemann-Cartan term, torsion terms, and vacuum energy. This reveals how each coupling constant modifies the dynamics.
\\
\\
The constancy equation now reads
\begin{equation}\label{Constancy equation}
\begin{split}
 g_1\left(R^{\alpha\beta\gamma\mu}R_{\alpha\beta\gamma}^{\nu}-\frac{1}{4}g^{\mu\nu}R^{\alpha\beta\gamma\xi}R_{\alpha\beta\gamma\xi}\right)\\-\frac{1}{8\pi G}\left(R^{(\mu\nu)}-\frac{1}{2}g^{\mu\nu}R+g^{\mu\nu}\Lambda_0\right)\\-2g_3\left( S^{\xi\alpha\mu}S_{\xi\alpha}^{\nu}-\frac{1}{2}S^{\mu\alpha\beta}S^{\nu}_{\alpha\beta}-\frac{1}{4}g^{\mu\nu}S_{\xi\alpha\beta}S^{\xi\alpha\beta}\right)=T^{(\mu\nu)}.
\end{split}
\end{equation}
Eqn.\eqref{Constancy equation} represents the fully generalized field equation of Einstein Cartan gravity with quadratic corrections, the core dynamical equation of the model.
\\
\\
Here the coupling constants $g_2$ and $g_4$ have been expressed in terms of the gravitational coupling constant $G$ and a constant vacuum energy term as:
\begin{equation} \label{1}
    g_1g_2=\frac{1}{16\pi G}=\frac{M_{Pl}^2}{2},
\end{equation}
\begin{equation}\label{2}
    6g_1g_2^2+g_4=\frac{\Lambda_0}{8\pi G}=M_p^2\Lambda_0,
\end{equation}
where $M_{Pl}=\sqrt{\frac{1}{8\pi G}}$ is the reduced Planck mass \cite{Baumann:2022cosmology,Mukhanov:2005fem,Weinberg:2008cosmology}.
\\
\\
Combining eqns.\eqref{1} and \eqref{2} yields
\begin{equation}
    \Lambda_0=3g_2+8\pi Gg_4=\frac{1}{M^2_{Pl}}\left(\frac{3M^2_{Pl}}{2g_1}+g_4\right).
\end{equation}
The parameter $g_1$ is the deformation parameter of the theory as it determines the relative strength of the quadratic Riemann-Cartan extension of Einstein's gravity. The coupling constant $g_2=\frac{M_p^2}{2g_1}$ is thus proportional to the inverse of that deformation parameter. Setting $\Lambda_0=0$ as per the zero energy Universe condition relates the coupling constants $g_1$ and $g_2$ to the vacuum energy density $g_4$:
\begin{equation}
    g_1=-\frac{3M_{Pl}^4}{2g_4}; g_2=-\frac{g_4}{3M_{Pl}^2}.
\end{equation}
These relations express the coupling constants $g_1,g_2,g_3,g_4$ in terms of Newton’s constant and the cosmological term, clarifying how $g_1$ acts as a deformation parameter measuring deviation from standard GR \cite{Smalley1978,Benisty2021,Shie_2008}.
\\
\\
With the general framework in place, we now turn to its cosmological realization by applying it to an FLRW Universe \cite{Baumann:2022cosmology,Mukhanov:2005fem,Weinberg:2008cosmology}.

\section{Torsion-Modified Cosmology in the DW Formalism} \label{Section 3}
To explore the cosmological consequences of the De Donder–Weyl (DW) formulation of Einstein–Cartan (EC) gravity, we now embed the theory into a homogeneous and isotropic background described by the Friedmann Lema\^{i}tre Robertson Walker ($FLRW$) metric \cite{Baumann:2022cosmology,Mukhanov:2005fem,Weinberg:2008cosmology}:
\begin{equation}
    ds^2=dt^2-a^2(t)\left[\frac{dr^2}{1-K_0r^2}+r^2(d\theta^2+sin^2\theta d\phi^2)\right].
\end{equation}
Substituting this metric into the generalized field eqn.\eqref{Constancy equation} yields the extended Friedmann equations derived in recent analyses of quadratic Einstein–Cartan and covariant canonical gauge gravity \cite{KirschVasakStruckmeier2023,vanDeVennVasakKirschStruckmeier2022}:
\begin{equation} \label{Friedmaneq 1}
\begin{split}
   3g_1\left[\frac{k^2}{a^4}+\frac{2k}{a^2}(H^2-s_0^2)-H^2(2\dot{H}+5s_0^2)-\dot{H}^2-2Hs_0\dot{s}_0+\dot{s}_0^2+s_0^4\right]\\+\rho_m+\rho_r+\frac{-3H^2+\Lambda_0+3s_0^2(1-8\pi Gg_3)}{8\pi G}+\frac{k}{8\pi Ga^2}= 0,
\end{split}
\end{equation}
\begin{equation}\label{Friedmanneq 2}
\begin{split}
  g_1\left[\frac{k^2}{a^4}+\frac{2k}{a^2}(H^2-s_0^2)-H^2(2\dot{H}+5s_0^2)-\dot{H}^2-2Hs_0\dot{s}_0+\dot{s}_0^2+s_0^4\right]\\+p_r+\frac{3H^2+2\dot{H}-\Lambda_0-3s_0^2(1-8\pi Gg_3)}{8\pi G}+\frac{k}{8\pi Ga^2}=0,
\end{split}
\end{equation}
the trace of eqn.\eqref{Friedmaneq 1} is found to be
\begin{equation}
\begin{split} \label{Trace of friemdann}
\rho_m+\frac{-3\dot{H}-6H^2+2\Lambda_0+3s_0^2(1-8\pi Gg_3)}{8\pi G}-\frac{3k}{4\pi Ga^2}= 0.
\end{split}
\end{equation}
Here $s_0$ is the timelike component of the vector $s^{\sigma}$ defined by \cite{Benisty2021,Shie_2008}
\begin{equation}
   s^{\sigma} =\frac{1}{\sqrt{3!}}\epsilon^{\alpha\,u\nu\sigma}S_{\alpha\mu\nu},
\end{equation}
which, owning to homogeneity and isotropy \cite{Hehl1976Foundations,Trautman1973,Smalley1978}, is given by
\begin{equation}
    s^{\sigma}=(s^0,0,0,0).
\end{equation}
Eqns.\eqref{Friedmaneq 1} and \eqref{Friedmanneq 2} represent the generalized Friedmann equations including quadratic Riemann Cartan corrections.
\\
\\
Following the $\Lambda CDM$ model, we introduce the cosmological parameters 
\begin{equation}
\Omega_m,\Omega_r,\Omega_\Lambda=\Lambda_0/3H_0^2,\Omega_K=-K_0/H_0^2,
\end{equation}
where the matter and radiation densities evolve as \cite{Baumann:2022cosmology,Mukhanov:2005fem,Weinberg:2008cosmology}
\begin{equation}
    \rho_m(a)=\rho_{cr}\Omega_ma^{-3},
\end{equation}
\begin{equation}
    \rho_r=\rho_{cr}\Omega_ra^{-4},
\end{equation}
where the critical density $\rho_{cr}$ is defined by 
\begin{equation}
    \rho_{cr}=\frac{3H_0^2}{8\pi G}=3M_p^2H_0^2.
\end{equation}
Using these definitions, eqns.\eqref{Friedmaneq 1} and \eqref{Friedmanneq 2} can be expressed in dimensionless form, convenient for numerical or analytic analysis \cite{Benisty2021}
\begin{equation}
    \dot{a}^2+V(a)=\Omega_k,
\end{equation}
\begin{equation}
    \ddot{a}a+\dot{a}^2-2M\dot{a}^2-\Omega_K +(\Omega_s-1)s^2a^2=0,
\end{equation}
where the auxiliary quantities are
\begin{subequations}
\begin{align}
M(a) &=\frac{1}{4}\Omega_ma^{-3}+\Omega_{\Lambda},\\
\Omega_g &=\frac{1}{32\pi GH_0^2g_1},\\
\Omega_s &= 8\pi Gg_3,\\
s &=\frac{s_0}{\sqrt{3!}H_0},\\
V(a)&=V_0(a)+V_{geo}(a)+V_{tor}(a),\\
V_0(a) &=-\Omega_ma^{-1}-\Omega_ra^{-2}-\Omega_{\Lambda}a^2,\\
V_{geo}(a,s) &=-\frac{M}{\Omega_g-M}\left(\frac{3}{4}\Omega_m a^{-1}+\Omega_ra^{-2}\right);
\end{align}
\end{subequations}
and 
\begin{equation}
\begin{split}
    V_{tor}=-\frac{1}{\Omega_g-M}\left[\begin{aligned}\frac{1}{4}(\dot{s}a+\dot{a}s)^2-s^2\dot{a}^2\\-\frac{\Omega_s}{2}s^2a^2\left(\frac{\dot{a}^2}{a^2}+\left(\frac{\Omega_s}{2}-1\right)s^2-\Omega_Ka^{-2}\right)\end{aligned}\right]\\+(\Omega_s-1)s^2a^2.
\end{split}
\end{equation}
Here, the dot derivative denotes derivative with respect to $\tau=tH_0$.
\\
\\
The potential functions $V_0,V_{geo},V_{tor}$ isolate, respectively, the standard cosmological contribution, the geometric correction due to the quadratic Riemann Cartan term, and the torsional correction associated with $g_3$ \cite{Benisty2021,Shie_2008}. This formulation reveals that torsion modifies the expansion history both directly, through $s^2$ explicit terms, and indirectly via its coupling to the curvature dependent potential $V_{geo}$.
\\
\\
In what follows, we shall assume that the Universe is flat and neglect the cosmological constant ($k=\Lambda_0=0$).
\\
\\
Now, after inflation, the Universe evolves through three major epochs: radiation domination (RDE), matter domination (MDE), and dark-energy domination ($\Lambda$DE) \cite{Baumann:2022cosmology,Mukhanov:2005fem,Weinberg:2008cosmology}. In the following, we analyze in detail the first two stages, focusing on how torsion and the quadratic Riemann Cartan corrections modify the standard power law evolution of the scale factor.
\\
\\
During the radiation era, the energy density is dominated by relativistic species with $\rho\propto a^{-4}$ and the cosmological constant and matter components can be neglected \cite{Baumann:2022cosmology,Mukhanov:2005fem,Weinberg:2008cosmology}. Assuming spatial flatness $K_0=0$, we adopt the commonly used power-law ansatz
\begin{equation}
    a=\beta t^{\alpha},
\end{equation}
where $H=\frac{\dot{a}}{a}$ and $\dot{H}=-\frac{\alpha}{t^2}$. Substituting this into eqn.\eqref{Trace of friemdann} yields an analytic form for the torsion amplitude $s_0$ \cite{Benisty2021,Shie_2008}:

\begin{equation} \label{torsion decay}
    s_0=\pm \sqrt{\frac{\alpha(2\alpha-1)}{1-8\pi Gg_3}}\frac{1}{t}.
\end{equation}
Eqn.\eqref{torsion decay} shows that torsion decays inversely with time, and vanishes in the limit $g_3\rightarrow \infty$, corresponding to torsionless GR.
\\
\\
Inserting this expression into the Friedmann equations gives \cite{vanDeVennVasakKirschStruckmeier2022}
\begin{equation}\label{f1}
    \frac{\rho_{r,0}}{\beta^4t^{4\alpha}}+\frac{C_1(\alpha)}{t^2}-g_1\frac{C_2(\alpha)}{t^4}=0,
\end{equation}
where
\begin{equation}
    C_1(\alpha)=\frac{3\alpha(\alpha-1)}{8\pi G},
\end{equation}
encodes the Einstein–Hilbert contribution and the quadratic torsion tensor $W^{\mu\nu}$, while 
\begin{equation}
\begin{aligned}
    C_2(\alpha)\\ =\frac{3\alpha(2\alpha-1)}{(1-8\pi Gg_3)^2}\left[(\alpha+1)(3\alpha+1)\begin{aligned}
    \\-8\pi G g_3(5\alpha^2-1)\\-64\pi^2G^2g_3^2\alpha\
    \end{aligned}
    \right],
    \end{aligned}
\end{equation}
captures the quadratic Riemann–Cartan correction.
\\
\\
Eqn.\eqref{f1} holds as a polynomial equation for a range of small times (where radiation domination persists). Hence, the respective coefficient for exponents of time have to vanish \footnote{See for example \cite{vanDeVennVasakKirschStruckmeier2022} for a more detailed treatment on the radiation dominated phase of such a setup.}. This leaves the admissible exponents
\begin{equation}
    \alpha\in\{1,\frac{1}{2}\}.
\end{equation}
For the case $\alpha=\frac{1}{2}$, we have $s_0=0$ and $C_2=0$. Furthermore,
\begin{equation}
    C_1\left(\frac{1}{2}\right)=\frac{-3}{32\pi G},
\end{equation}
and thus, by eqn.\eqref{f1}; we have
\begin{equation}
    \beta^4=\frac{32\pi G\rho_{r,0}}{3}.
\end{equation}
Hence, in the $\alpha=\frac{1}{2}$ case, the expansion of the Universe is given by that standard $\Lambda$ CDM expression
\begin{equation}
    a=\left(\frac{32\pi G}{3}\rho_{r,0}\right)^{1/4}\sqrt{t}.
\end{equation}
For the last case $\alpha=1$ we have
\begin{equation}
    s_0=\pm \frac{1}{\sqrt{1-8\pi Gg_3}}\frac{1}{t},
\end{equation}
\begin{equation}
C_1(1)=0,
\end{equation}
\begin{equation}
    C_2(1)=\frac{24-96\pi Gg_3(1+2\pi G g_3)}{(1-8\pi Gg_3)^2}.
\end{equation}
Finally, we get 
\begin{equation}
    \beta^4=\frac{\rho_{r,0}}{g_1C_2(1)},
\end{equation}
and
\begin{equation}\label{Miline}
    a=\left(\frac{\rho_{r,0}}{g_1C_2(1)}\right)^{1/4}t.
\end{equation}
These two branches illustrate how the power-law assumption yields discrete expansion behaviors rather than a continuous family of solutions. Only one branch ($\alpha=\frac{1}{2}$) reproduces the observed early-Universe scaling, while the second produces nonstandard cosmologies inconsistent with current constraints on radiation-era expansion and big-bang nucleosynthesis (BBN) \cite{Fields2023,Grohs2023,PDG2024,Tytler2000,LiDodelsonHu2019,KamionkowskiSpergelSugiyama1994}.
\\
\\
During matter domination, non relativistic matter dominates the energy density ($\rho_m\propto a^{-3}$), while radiation and $\Lambda$ can be neglected \cite{Baumann:2022cosmology,Mukhanov:2005fem,Weinberg:2008cosmology}. Using the same ansatz $\alpha=\beta t^{\alpha}$ and assuming $K_0=0$, eqn.\eqref{Trace of friemdann} yields
\begin{equation}
      s_0=\pm \sqrt{\frac{\alpha(2\alpha-1)-4\pi\beta^{-3} G\rho_{m,0}t^{2-3\alpha}}{1-8\pi Gg_3}}\frac{1}{t},
\end{equation}
showing that torsion now depends on the evolving matter density. Substituting into the Friedmann equation leads to
\begin{equation}
\begin{split}\label{10}
\rho_{m,0}\beta^{-3}t^{-3\alpha}+\frac{3[\alpha^2-\alpha -4\pi G \beta^{-3}\rho_{m,0}t^{2-3\alpha}]}{8\pi G}\frac{1}{t^2}\\+g_1\left[B(\alpha,t,g_3)
       \right]=0,
    \end{split}
\end{equation} 
where,
\begin{equation}
\begin{split}
 B(\alpha,t,g_3)=-\frac{\alpha^2}{t^2}\left(-2\alpha\frac{1}{t^2}+\frac{5}{t^2}\left(\frac{\alpha(2\alpha-1)-4\pi\beta^{-3} G\rho_{m,0}t^{2-3\alpha}}{1-8\pi Gg_3}\right)\right)\\+\frac{\alpha^2}{t^4}-2H\frac{\alpha(2\alpha-1)-4\pi\beta^{-3} G\rho_{m,0}t^{2-3\alpha}}{1-8\pi Gg_3}\frac{1}{t^2}\\-\frac{4H\pi\beta^{-3}(2-3\alpha)G\rho_{m,0}t^{-3\alpha}}{1-8\pi Gg_3}\\+\left( 
 \begin{aligned}
 -\sqrt{\frac{\alpha(2\alpha-1)-4\pi\beta^{-3} G\rho_{m,0}t^{2-3\alpha}}{1-8\pi Gg_3}}\frac{1}{t^2}\\-\frac{2\pi\beta^{-3} G \rho_{m,0}(2-3\alpha)t^{-3\alpha}}{\sqrt{(1-8\pi G g_3)(\alpha(2\alpha-1)-4\pi\beta G\rho_{m,0}t^{2-3\alpha})}}
 \end{aligned}
 \right)^2\\
+\left(\frac{\alpha(2\alpha-1)-4\pi\beta^{-3} G\rho_{m,0}t^{2-3\alpha}}{1-8\pi Gg_3}\right)^2\frac{1}{t^4}.
\end{split}
\end{equation}
Requiring that coefficients of different powers of $t$ vanish simultaneously yields the consistent exponent
\begin{equation}
    \alpha=\frac{2}{3},
\end{equation}
which reproduces the standard $\Lambda CDM$ matter domination era scaling $a\propto t^{3/2}$.
\\
\\
Using eqn.\eqref{10} then gives us
\begin{equation} \label{Matter}
\begin{split}
\frac{-1}{g_1}\left(\rho_{m,0}\beta^{-3}t^{-2}+\frac{3\left[\frac{2}{9}-4\pi G \beta^{-3}\rho_{m,0}\right]}{8\pi G}t^{-2}\right)\\
=-\frac{4}{9t^2}\left(-\frac{4}{3}\frac{1}{t^2}+\frac{5}{t^2}\left(\frac{\frac{2}{9}-4\pi\beta^{-3} G\rho_{m,0}}{1-8\pi Gg_3}\right)\right)\\-\frac{4}{9t^4}-\frac{4}{3}\frac{\frac{2}{9}-4\pi\beta^{-3} G\rho_{m,0}}{1-8\pi Gg_3}\frac{1}{t^3}\\+\frac{\frac{4}{9}-4\pi\beta^{-3} G\rho_{m,0}}{1-8\pi Gg_3}\frac{1}{t^4}\\+\left(\frac{\frac{4}{9}-4\pi\beta^{-3} G\rho_{m,0}}{1-8\pi Gg_3}\right)^2\frac{1}{t^4}.
\end{split}
\end{equation} 
From which, we infer
\begin{equation}
    \beta=\left(6\pi G \rho_{m,0}\right)^{1/3}.
\end{equation}
This means that in the case $\alpha=\frac{2}{3}$, the expansion of the Universe is described by
\begin{equation}
    a= \left(6\pi G \rho_{m,0}\right)^{1/3}t^{2/3},
\end{equation}
which corresponds to the standard $\Lambda CDM$ model.
\\
\\
However, the torsion sector still imposes only a discrete set of allowed values of$g_3$:-
\begin{equation}
    g_3=\frac{1}{8\pi G}\left[1-\frac{8}{-62+48t\pm \sqrt{(62-48t)^2-192}}\right],
\end{equation}
together with the torsionless branch $g_3=\infty$.
\\
\\
This particular highly discrete behaviour is rather unphysical. Furthermore, dependence on $t$ is also unphysical, physical coupling constants such as $g_3$ cannot depend on cosmic time.
\\
\\
This means that the only consistent physical value in this branch is the torsionless limit $g_3\rightarrow \infty$, reducing the theory back to standard GR.
\\
\\
A further, even more pathological branch emerges when $g_1\not= 0$ but $g_3\rightarrow \infty$, yielding a solution with
\begin{equation}
    \alpha=\frac{4}{3}.
\end{equation}
The Friedmann equation \eqref{10} now reads
\begin{equation}
\begin{split}
 \rho_{m,0}\beta^{-3}t^{-2}+\frac{-\frac{4}{3}-12\pi G \beta^{-3}\rho_{m,0}t^{-2}}{8\pi G}t^{-2}+g_1\frac{20}{27t^4}=0.
\end{split}
\end{equation}
The corresponding normalization is
\begin{equation}
    \beta=(-6\pi G\rho_{m,0})^{1/3},
\end{equation}
with $g_1$ having only $1$ allowed value\footnote{Such a regime corresponds to a purely quadratic–geometric phase where the Riemann–Cartan corrections dominate the expansion rate. Similar high $g_1$ behaviors have been observed in numerical studies of torsion driven cosmologies \cite{Benisty2021,Shie_2008}.},
\begin{equation}
    g_1=\frac{1}{5}\left(\frac{3}{4}\right)^3.
\end{equation}
The scale factor now becomes
\begin{equation}
    a(t)=(-6\pi G\rho_{m,0})^{1/3}t^{4/3},
\end{equation}
which is negative for all $t>0$.
\\
\\
Since the scale factor represents a physical spatial volume, 
\begin{equation}
    a(t)>0,
\end{equation}
is a fundamental requirement. A negative $a(t)$ has no interpretation in FLRW cosmology and therefore signals an unphysical branch generated solely by the restrictions of the power-law ansatz.
\\
\\
To conclude, the matter-dominated sector exhibits:-
\begin{itemize}
    \item Discrete allowed values of $g_3$ instead of a continuous parameter space.
    \item Explicit time dependence in $g_3$ unless one chooses the trivial torsionless limit.
    \item A branch with a negative scale factor, which is physically meaningless.
    \item Sensitivity to $g_1$ that switches the Universe between qualitatively different expansion laws, which is not seen in realistic cosmology.
\end{itemize}
All these issues arise because the power-law ansatz forces the Friedmann system to collapse onto a handful of isolated algebraic conditions. Hence, the power law anstaz must be rejected in the matter dominated case. 
\\
\\
One can also see that you get a completely different behavior of physical observables which do not coincide with their observed values even on the radiation dominated epoch. For example, one arrives at a completely different physics than the $\Lambda CDM$ model when one studies big bang nucleosynthesesis \cite{Cyburt2016,Fields2023,Grohs2023,PDG2024,Tytler2000} like we shall show below. 
\\
\\
Let us consider the particular case of $\alpha =1$ in the radiation dominated epoch. Then, $H=\frac{1}{t}=\frac{1}{a}\left(\frac{\rho_{r,0}}{g_1C_2(1)}\right)^{1/4}\propto T\left(\frac{\rho_{r,0}}{g_1C_2(1)}\right)^{1/4}.$
 \\
 \\
Recall that by statistical mechanics, we have \cite{Baumann:2022cosmology,Mukhanov:2005fem,Weinberg:2008cosmology}
\begin{equation}\label{Stat.mech}
    \rho=\rho_{r,0}a^{-4}=\frac{\pi ^2}{30}g_{*}T^{4},
\end{equation}
during BBN \cite{Cyburt2016,Fields2023,Grohs2023,PDG2024,Tytler2000}. Using eqn.\eqref{Stat.mech}, we find
\begin{equation} \label{temperature 1}
    a^{-1}=\left(\frac{\pi ^2g_{*}}{30\rho_{r,0}}\right)^{1/4}T.
\end{equation}
This means that 
\begin{equation}
    H=\left(\frac{\pi ^2 g_{*}}{30g_1C_2(1)}\right)^{1/4}T.
\end{equation}
The normalized relic abundance of a particle species $X$ is then
\begin{equation}
\begin{split}
    \Omega_{X}=\frac{\rho_{X,0}}{\rho_{crit,0}}\\=\frac{H(M_{X})}{M^2_X}\frac{T_0^3}{3M_{PI}^2H_0^2}\frac{x_f}{<\sigma v>}\frac{g_{*S}(T)}{g_{*S}(M_X)}\\=\left(\frac{\pi ^2 g_{*}}{30g_1C_2(1)}\right)^{1/4}\frac{T_0^3}{3M_{PI}^2M_{X}H_0^2}\frac{x_f}{<\sigma v>}\frac{g_{*S}(T)}{g_{*S}(M_X)},
\end{split}
\end{equation}
which numerically gives
\begin{equation}
\begin{split}
    \Omega_{X}\sim \\
    0.2 \frac{1}{(g_{*}(M_X))^{3/4}(g_2C_2(1))^{1/4}}\frac{M_{Pl}}{M_X}\frac{10^{-8}GeV^{-2}}{<\sigma v>}\sim \\
    0.2 \frac{1}{(g_{*}(M_X))^{3/4}(g_2C_2(1))^{1/4}}\frac{M_{Pl}}{M_X}\frac{10^{-8}GeV^{-2}}{<\sigma v>}.
\end{split}
\end{equation}
We immediately notice a peculiar $\propto \frac{1}{M_X}$ scaling which is completely different form the prediction of standard $\Lambda CDM $ cosmology. Furthermore, $\Omega_{X}$ is proportional to $\frac{1}{g_1^{1/4}}$ and hence would quite larger than the observed value.
\\
\\
The radiation- and matter-era analyses show that imposing the power-law ansatz 
\begin{equation}
    a(t)=\beta t^{\alpha}
\end{equation}
forces the Hubble parameter into the universal form
\begin{equation}
    H=\frac{\dot{a}}{a}=\frac{\alpha}{t}\propto \frac{1}{t}
\end{equation}
This behavior for standard $\Lambda CDM$, but becomes excessively restrictive in the torsionful Einstein Cartan DW system. In particular, when $g_1\not=0$ the quadratic Riemann Cartan term generates nonlinear couplings between $a(t),\dot{a}(t)$, and the torsion amplitude $s_0(t)$. These nonlinearities introduce explicit time dependence in the effective expansion rate that cannot be absorbed into the simple form $H\propto 1/t$. This is precisely why the power-law ansatz yields only a handful of discrete and often unphysical branches.
\\
\\
To reveal the true structure of the dynamics, we therefore turn to the special case $g_1=0$. When the quadratic Riemann–Cartan term is removed, the equations become analytically tractable, and a more flexible hybrid solution naturally emerges:
\begin{equation}
a(t)=Ct^{\alpha}exp(Kt^{\beta})
\end{equation}
which no longer enforces the $H\propto 1/t$ behavior. This solution captures the genuine torsion-driven modification to the expansion history.
\\
\\
In the radition dominated era, the above ansatz turn into
\begin{equation}
    a(t)=Ct^{1/2}e^{-2A\sqrt{t}},
\end{equation}
while, in the matter dominated era, we have
\begin{equation}
    a(t)=C t^{2/3} e^{3Bt^{1/3}}.
\end{equation}
\\
\\
Here,
\begin{equation}
    A(s_0)=\frac{3 - \sqrt{9+32ts_0^2(1-8\pi Gg_3)}}{8\sqrt{t}},
\end{equation}
and $B(s_0)$ satisfies
\begin{equation}
  \frac{2} {9}t^{-2}(e^{-9Bt^{1/3}}-1)-2Bt^{-5/3}-2B^2t^{-4/3}+s_0^2(1-8\pi Gg_3)=0.
\end{equation}
Note that $A$ and $B$ both vanish in the usual $\Lambda CDM$ model \cite{Baumann:2022cosmology,Mukhanov:2005fem,Weinberg:2008cosmology}. Now, $A$ and $B$ both should be time independent according to the assumptions we made to arrive at this solution. This means $\dot{A}$ and $\dot{B}$ both vanish at all times. This in turn constraints $s_0$ to satisfy a differential equation. For the radiation dominated phase, that differential equation is given by
\begin{equation}
    -3\sqrt{9+8t^2s_0^2(1-8\pi Gg_3)}+9-8t^2s_0^2(s_0+2\dot{s_0}t)(1-8\pi Gg_3)=0.
\end{equation}
While, for the matter dominated phase, the differential equation is given by
\begin{equation}
\begin{split}
  -\frac{8}{9}t^{-3}(e^{-9Bt^{1/3}}+1)+\frac{2}{3}Bt^{-8/3}(5-e^{-9Bt^{1/3}})-\frac{8}{3}B^2t^{-7/3}\\+2s_0\dot{s}_0(1-8\pi Gg_3)=0.
\end{split}
\end{equation}
Analytically solving the Friedmann equations in this model in a general case ($g_1\not=0$) is almost impossible (apart from the trivial solutions $a=\beta t^{\alpha}$ we find in the $\kappa=0$ case) due to the non-linearity of $s_0$ which appears in the term coupled to $g_1$. 
\\
\\
To connect the model with observation, consider the temperature evolution during the radiation era, derived from eqn.\eqref{temperature 1}: 
\begin{equation}
    T=\left(\frac{90}{32\pi^3 Gg_{*}}\right)^{1/4}t^{-1/2}e^{-2A\sqrt{t}}.
\end{equation}
At the surface of last scattering \cite{LiDodelsonHu2019,KamionkowskiSpergelSugiyama1994,Bolejko2011,Planck2015XVIII2016}, corresponding to $t\simeq 3.6 \times 10^5 yrs$ to $t\simeq 4.0 \times 10^{5} yrs$, we equate the temperature predicted by the $CMB$ \cite{Durrer2015_CMBHistory,Bucher2015_CMBAnisotropy,KamionkowskiKosowsky1999_CMBParticlePhysics,Jungman1996_CosmologicalParameterCMB,SunyaevKhatri2013_CMBDistortions} to estimate bounds on $A$:
\begin{equation}
   0> A>-6.5\times 10^{-4}(yr)^{-1/2}.
\end{equation}
The torsion $s_0$ corresponding to $A=6.5\times 10^{-4}yr^{-1/2}$ at $t=400,000yrs$ is then
\begin{equation}
    s_0=1.19\times 10^{-11}(yr)^{-1}.
\end{equation}
These limits indicate that any residual torsion today must be exceedingly small, though non-zero torsion in the early Universe could still produce measurable imprints on cosmological observables.
\section{Conclusions and Outlook} \label{Section 4}
In this work, we have investigated the cosmological implications of Einstein Cartan gravity formulated within the De Donder-Weyl Hamiltonian framework \cite{Weyl1935,Caratheodory1929,GoldschmidtSternberg1973,Kastrup1983,Kanatchikov2000,Gotay1998,Forger2005,Struckmeier2008,Struckmeier2015,Vasak2020,KirschVasakStruckmeier2023,vanDeVennVasakKirschStruckmeier2022}, where the tetrad and Lorentz connection are treated as independent dynamical variables \cite{Trautman1972,Shapiro2002,Capozziello2015}. Embedding this formulation into a homogeneous and isotropic FLRW background \cite{Baumann:2022cosmology,Mukhanov:2005fem,Weinberg:2008cosmology} led to a set of torsion-modified Friedmann equations that exhibit rich and nontrivial dynamics across cosmic epochs \cite{Smalley1978,Vasak2020,Benisty2021,Shie_2008}.
\\
\\
We began by revisiting the radiation-dominated era and then extended the analysis to the matter dominated epoch, applying the same power law ansatz $a(t)=\beta t^{\alpha}$ \cite{Benisty2021,Shie_2008}. In the radiation era, this ansatz admits only two possible branches,$\alpha=\frac{1}{2}$ and $\alpha=1$, of which only the first coincides with the standard $\Lambda CDM$ behaviour \cite{Baumann:2022cosmology,Mukhanov:2005fem,Weinberg:2008cosmology}. The second branch fails to reproduce the expected radiation scaling and leads to modified temperature-time relations incompatible with BBN. This already indicates that the power-law ansatz introduces artificially discrete and partially unphysical behavior in the torsion-modified equations.
\\
\\
In the matter-dominated epoch this discreteness becomes even more pronounced: the consistency conditions restrict the coupling $g_3$ to isolated discrete values, and in the non-standard branches even introduce an explicit time dependence in the effective value of $g_3$, which is physically unacceptable.
\\
\\
Moreover, an additional branch with $\alpha=\frac{4}{3}$ appears in the torsion less limit $g_3\rightarrow \infty $, where the solution requires
\begin{equation}
    \beta=(-6\pi G\rho_{m,0})^{1/3}
\end{equation}
implying a negative scale factor, which is manifestly unphysical. These effects show that the power-law ansatz does not correctly capture the nonlinear torsion–geometry couplings of the DW Einstein-Cartan system and instead forces the model into highly discretized and physically problematic branches.
\\
\\
For $g_1=0$, where the quadratic Riemann-Cartan term vanishes, we identified instead a hybrid solution
\begin{equation}
    a(t)=Ct^{\alpha}e^{Kt^{\beta}}.
\end{equation}
When $g_{1}\not = 0$, the Friedmann equations become nonlinear, precluding closed-form analytic solutions; however, torsion-induced terms then contribute higher-order corrections that significantly influence early-Universe dynamics \cite{Benisty2021,Shie_2008,Vasak2020,Cyburt2016}.
\\
\\
It is worth emphasizing that inflation was not included in this study. In torsionful cosmologies, the intrinsic spin–torsion coupling naturally avoids the initial singularity and provides an alternative mechanism for early accelerated expansion, leading instead to a nonsingular Big Bounce model \cite{Cartan1922,Cartan1923,Cartan1924,Sciama1964a,Kibble1961,Trautman1972,Hehl1971,Trautman1973,Hehl1979_book,HehlObukhov2007,Blagojevic2013,Capozziello2015,Smalley1978}. In this picture, cosmic acceleration emerges as a geometric consequence of torsion rather than as an effect of an external inflaton field \cite{Benisty2021,Shie_2008,Vasak2020,Cyburt2016,Bolejko2011,Planck2015XVIII2016}.
\\
\\
Looking ahead, a comprehensive numerical analysis of the full torsion-modified Friedmann system is essential to explore the nonlinear regime and the parameter space of $g_1$ and $g_3$. In future works, we plan to build upon the current work and study torsion-driven alternatives to dynamical dark energy \cite{Bolejko_2020,vanDeVennVasakKirschStruckmeier2022,Benisty2021,Shie_2008}. Exploring this regime could reveal new signatures of torsion in the late-time Universe, potentially linking geometric effects to dark-energy phenomenology and observational probes such as the CMB and large-scale structure \cite{Tytler2000,LiDodelsonHu2019,Bolejko2011,KamionkowskiSpergelSugiyama1994,Planck2015XVIII2016,Durrer2015_CMBHistory,Bucher2015_CMBAnisotropy,KamionkowskiKosowsky1999_CMBParticlePhysics,Jungman1996_CosmologicalParameterCMB,SunyaevKhatri2013_CMBDistortions}. 
\section{Acknowledgments}
We express our sincere gratitude to Oem Trivedi for his guidance, insightful discussions, and constant support throughout the development of this work. We also extend our heartfelt thanks to Aum Trivedi for his encouragement and helpful suggestions during various stages of this project. We are likewise grateful to Meet Vyas for his valuable academic inputs, and to Kanabar Jay for his collaborative assistance and constructive feedback during the preparation of this manuscript. The work of M.K. was performed in Southern Federal University with financial support of grant of Russian Science Foundation № 25-07-IF.

\bibliographystyle{spphys}
\bibliography{references}

\end{document}